\documentclass[a4paper,11pt]{article}
\usepackage{jcappub}
\usepackage{amssymb}
\usepackage{graphicx}
\usepackage{amsmath}
\usepackage{hyperref}

\title{\boldmath Primordial black holes spin from cosmological first-order phase transitions}

\author[1]{Yu-Shi Hao,}

\emailAdd{haoys@mail.tsinghua.edu.cn}
\affiliation[1]{Department of Physics, Tsinghua University, Beijing,100084, China}

\abstract{The stochastic bubble nucleation during cosmological first-order phase transitions leads to variations in the phase transition initiation times across different Hubble volumes, thereby generating non-Gaussian density perturbations in regions with delayed transitions. Based on the accumulation mechanism and the false vacuum island model . This paper investigates the spin angular momentum of primordial black holes formed from nonspherical collapse. By introducing the nucleation history integration method, without assuming a Gaussian distribution, we calculate the expectation values and variances of the semi-axis lengths of overdense ellipsoidal regions, combined with the statistical properties of the velocity shear tensor, we derive the quantitative relationship between the Kerr parameter $a_*$ describing black hole spin and the phase transition parameters , latent heat strength $\alpha$ and phase transition rate $\beta$. The study finds that the Kerr parameter increases with $\alpha$ and decreases with $\beta$; estimate the typical the magnitude of $a_*$ can reach $10^{-3}$, which is significantly higher than that of primordial black holes formed in the radiation-dominated era under peak theory, but still lower than that in a matter-dominated era.}

\begin{document}
\maketitle
\flushbottom

\section{Introduction}\label{sec:introduction}

The cosmological first-order phase transition (FOPT) ~\cite{Mazumdar:2018dfl,Hindmarsh:2020hop,Caldwell:2022qsj,Athron:2023xlk} is a stochastic process in the early universe where a metastable false vacuum high-energy state tunnels into a stable true vacuum low-energy state. 
The probability of this transition can be described by instanton and saddle-point solutions. Cosmological first-order phase transitions play a crucial role in probing new physics~\cite{Cai:2017cbj,Bian:2021ini} beyond the Standard Model of particle physics in the early universe, with key observational prospects coming from the detection of stochastic gravitational wave backgrounds~\cite{Caprini:2015zlo,Caprini:2019egz,LISACosmologyWorkingGroup:2022jok} by space-based detectors such as LISA~\cite{Armano:2016bkm,Audley:2017drz,LISA:2022yao}, Taiji~\cite{Hu:2017mde,Ruan:2018tsw,TaijiScientific:2021qgx} as well as TianQin~\cite{TianQin:2015yph,Luo:2020bls,TianQin:2020hid}.
Furthermore, the non-equilibrium conditions provided by FOPTs offer theoretical support for baryogenesis mechanisms (fulfilling the Sakharov conditions) to explain the matter-antimatter asymmetry , and provides theoretical support for the generation of primordial magnetic fields~\cite{Di:2020kbw}.

Primordial black holes (PBHs)~\cite{Shibata:1999zs,Niemeyer:1997mt,Polnarev:2006aa,Musco:2004ak,Musco:2008hv} are formed in the early universe from the gravitational collapse of localized overdense regions caused by density perturbations. Their mass range far exceeds that of black holes formed from stellar death, spanning from microscopic scales to supermassive scales, making them a compelling candidate for dark matter~\cite{Chapline:1975ojl}. Among various formation mechanisms during FOPT, in addition to the colliding mechanism proposed in the early stages~\cite{Hawking:1982ga,Crawford:1982yz,Moss:1994pi,Moss:1994iq} , recent studies have suggested the shrinking mechanism~\cite{Baker:2021nyl,Kawana:2021tde} and the accumulation mechanism~\cite{Liu:2021svg,Cai:2024nln} on which our research is based.

Our study is based on the primordial black hole formation mechanism known as the accumulating mechanism~\cite{Liu:2021svg,Cai:2024nln}, which is model-independent and imposes no special requirements on the particle properties of the phase transition model. This mechanism relies only on the randomness of vacuum bubble nucleation during the phase transition process. This randomness leads to  asynchronous phase transition histories in different regions of the universe,  causing an inhomogeneous evolution of the energy density. 
If a region begins vacuum bubble nucleation later than the surrounding background, it will possess a higher energy density. When such an overdense region re-enters the cosmological horizon, a primordial black hole will form if the density perturbation exceeds the critical threshold.
The threshold for the formation of primordial black holes has been extensively discussed through analytical methods and numerical simulations~\cite{Polnarev:2006aa,Harada:2013epa}.
These delayed-transition regions are similar to the Trapped False Vacuum Domains (TFVD) mechanism proposed earlier~\cite{Sato:1981bf,Maeda:1981gw,Sato:1981gv,Kodama:1981gu,Sato:1981hk,Kodama:1982sf}. Building upon this delayed phase transition mechanism, subsequent research has achieved typical model constructions for primordial black hole formation at the QCD 
scale~\cite{He:2022amv,He:2023ado,Gouttenoire:2023bqy,Salvio:2023blb}, the electroweak scale~\cite{He:2022amv,He:2023ado,Gouttenoire:2023bqy,Salvio:2023blb}, and the PeV scale~\cite{Baldes:2023rqv}.
Within the framework of this delayed phase transition mechanism, recent studies~\cite{Kawana:2022olo,Gouttenoire:2023naa,Lewicki:2023ioy} have semi-analytically or numerically investigated the dependence of the mass and abundance of the primordial black holes on the first-order phase transition parameters. 

Notably, primordial black holes can form in regions that remain entirely in the false vacuum, i.e., regions where no vacuum bubble has ever nucleated with a quadratic correction phase transition tunneling rate~\cite{Lewicki:2023ioy,Kanemura:2024pae}. In this FOPT model, vacuum bubble nucleation is highly concentrated around the peak nucleation time, followed by a sharp decline in the nucleation rate, hence this process can be regarded as an instantaneous phase transition.

The mass and abundance of PBHs produced from cosmological FOPT have been extensively studied. However, the spin (angular momentum) of PBHs also has a significant observational value, influencing their Hawking evaporation spectrum~\cite{Dasgupta:2019cae}and triggering superradiant instabilities~\cite{Calza:2023rjt}. Existing studies on PBH spin often operate within the peak theory framework under Gaussian density perturbations~\cite{Bardeen:1985tr,DeLuca:2019buf,Harada:2020pzb,Saito:2023fpt}. However, the density perturbations generated during FOPT are inherently non-Gaussian. Therefore, we employ the nucleation history integration method~\cite{Cai:2024nln}
to investigate the angular momentum of PBHs produced from first-order phase transitions.

In Chapter 2, we introduce the origin of primordial black hole angular momentum, demonstrating that during non-spherical collapse, the first-order angular momentum arises from differences in the principal axes length of the ellipsoid within the overdense region. Chapter 3 discusses the density perturbations generated by the stochastic cosmological first-order phase transition process. In Chapter 4, we analyze the relationship between phase transition parameters and the Kerr parameter characterizing black hole spin, comparing the primordial black hole angular momentum derived from peak theory under Gaussian assumptions versus that from the nucleation processes integration theory without requiring Gaussian assumptions. The final chapter presents conclusions and summarizes key findings.

\section{Primordial black hole spin}
This chapter discusses the definition of primordial black hole spin and its first-order contributions ( Second-order description of the PBH spin can be followed in ~\cite{DeLuca:2019buf}). By expanding density perturbations to the second order, we obtain an ellipsoidal overdense region. The deviation of the ellipsoidal semi-axes length from the horizon scale at the time of primordial black hole formation leads to non-spherical collapse, thereby generating spin.

The definition of primordial black hole spin is followed
in Ref.~\cite{DeLuca:2019buf,Harada:2020pzb,Saito:2023fpt}, and our work adopts their formulation.

We consider a flat FLRW metric,
\begin{align}
ds^2 = a^2(-d\eta^2 + dx^2 + dy^2 + dz^2) ,
\end{align}
where $a$ is the cosmological scale factor and $\eta$ is conformal time,assuming that the cosmic content is an ideal fluid with an energy-momentum tensor satisfying
\begin{align}
T^{ab} = \rho u^a u^b + p(g^{ab} + u^a u^b) .
\end{align}

For a region $\Sigma$ on spacelike hypersurface, the spin can be defined via a rotational Killing vector as
\begin{align}
S_i = \frac{1}{16\pi G} \int_{\partial \Sigma} \epsilon_{abcd} \nabla^c (\phi_i)^d
= -\frac{1}{8\pi G} \int_\Sigma R^{ab} n_a (\phi_i)_b d\Sigma
= -\int_\Sigma T^{ab} n_a (\phi_i)_b d\Sigma ,
\end{align}
where $n_a$ is the unit normal vector to the hypersurface, and the rotational Killing vectors are given by
\begin{align}
(\phi_i)^a = \epsilon_{ijk} (x - x_{pk})^j \delta^{kl} \left( \frac{\partial}{\partial x^l} \right)^a .
\end{align}

Finally, the spin is given by
\begin{align}
S_i = \frac{4}{3} a^4\epsilon_{ijk} \int_\Sigma d^3x \rho (x - x_{pk})_j (v - v_{pk})_k .
\end{align}
 where $x$ is the comoving coordinate.

For the process of collapse of an overdense region into primordial black holes, we consider the boundary of the integration region as an isodensity perturbation surface $\Sigma = \left\{ \mathbf{x} \mid \delta(\mathbf{x}) > \delta \right\}$. We expand the density perturbation at the density peak to second order,
\begin{align}\delta=\delta_{pk}+\frac{1}{2}\xi_{ij}(x-x_{pk})_i(x-x_{pk})_j ,
\end{align}
at the density perturbation peak, the first order derivative is zero, so only the second derivative contributes,
\begin{align}\xi_{ij}=\frac{\partial^2\delta}{\partial x^i\partial x^j} ,
\end{align}

Within the framework of the second-order expansion, the integration region is described by a quadratic form. By choosing a suitable coordinate system to diagonalize this quadratic form, we obtain the density field expression 
\begin{align}\delta \approx\delta_{pk}+\frac{1}{2}\sum^{i=3}_{i=1}\lambda_i(x-x_{pk})^2_i ,
\end{align}
where $\lambda_i$ are the second derivatives of the density perturbation along the three principal axes. We obtain an elliptical collapse region. Since the expansion is truncated at second order, this region maintains ellipsoidal symmetry. Therefore, only three degrees of freedom,the lengths of its principal axes, or equivalently, the three eigenvalues $\lambda_i$ are required to describe its shape. In this simplified model, we characterize the non-spherical symmetry of the collapse using the characteristic lengths (or the statistical expectation values of the eigenvalue) of these principal axes. It should be noted that the actual nonlinear collapse process can involve more complex, non-ellipsoidal morphologies.
 The ellipse equation is
\begin{align}\frac{(x-x_{pk})^2_1}{\frac{2(\delta-\delta_{pk})}{\lambda_1}}+\frac{(x-x_{pk})^2_2}{\frac{2(\delta-\delta_{pk})}{\lambda_2}}+\frac{(x-x_{pk})^2_3}{\frac{2(\delta-\delta_{pk})}{\lambda_3}}=1 ,
\end{align}
the comoving semi-axes are
\begin{align}\label{eq:2.10}a_i=\sqrt{\frac{2(\delta-\delta_{pk})}{\lambda_i}} ,
\end{align}
to characterize the local shape of the overdensity region, we expand the density perturbation $\delta$ around the peak position and retain terms up to the second order, neglecting all derivatives of the third and higher order. In the principal-axis frame of the Hessian matrix, the density contrast decreases away from its peak value along the three orthogonal directions $i=x,y,z$, which label the principal axes. The corresponding deviations from the peak value determine the characteristic extents of the overdense region along these directions, leading to the eigenvalues 
\begin{align}\label{eq:2.11}-\lambda_i=2a^2H^2(\delta_{pk}-\delta_i(t)) ,
\end{align}
 $\delta$ is the perturbation value on the isodensity surface at the boundary of the collapse region.In our previous work, this perturbation was defined as $\delta(t_0)$, corresponding to the nucleation region at initial time $t_0$. The quantities $\delta_i(t)$ are the local density perturbations outside the comoving 
Hubble radius $R = 1/(aH)$ measured along each of the three principal axes; they arise 
randomly from the phase transition dynamics. We convert the non-spherical symmetry of the collapse region into random values of density perturbations. When $\delta_i(t) = \delta(t_0)$, the semi-axis $a_i$ returns to the radius of the spherically symmetric collapse region $R = 1/aH$. When we introduce a perturbation $\delta(t_0 + \Delta t)$ on top of $\delta(t_0)$, the semi-axes deviate from $R = 1/aH$, which is the origin of non-spherical symmetry in gravitational collapse.

We consider the first-order part of the angular momentum.In the first-order calculation of angular momentum, we neglect the density distribution within a Hubble region.Therefore, $\rho$ is the average density of the overdense region.
\begin{align}\label{eq:2.12}S_i=\frac{4}{3}a^4\rho\epsilon_{ijk}v_{kl}\int d^3x x_jx_l,
\end{align}
where the velocity shear is
\begin{align}v_{kl}=\frac{\partial v_k}{\partial x_l},
\end{align}
to compute~\eqref{eq:2.12},we integrate
\begin{align}\int_\Sigma d^3x=a_1a_2a_3\int_0^1r^2dr\int_0^{2\pi}d\phi\int_0^{\pi}d\theta \sin{\theta},
\end{align}
we identifying
\begin{align}x_1=a_1r\cos{\phi}\sin{\theta},x_2=a_2r\sin{\phi}\sin{\theta},x_3=a_3r\cos{\theta},
\end{align}
then we can get the total angular momentum 
\begin{align}\label{eq:2.16}S=\frac{16\pi}{45}a^4\rho a_1a_2a_3(v_{23}[a_2^2-a_3^2],v_{13}[a_3^2-a_1^2],v_{12}[a_1^2-a_2^2]).
\end{align}
For example the angular momentum along the x-axis is proportional to the difference in the squares of the semi-axes along the y-axis and z-axis, \begin{align}S_1=\frac{4}{3}a^4\rho\int d^3x (v_{32}x_2x_2-v_{23}x_3x_3)=\frac{16\pi}{45}a^4\rho a_1a_2a_3v_{23}[a_2^2-a_3^2].
\end{align}

The semi-axes $a_i$ are inversely proportional to $aH$, so the semi-axes in $S$ contribute $a^{-5}$. The velocity shear is proportional to $a$, and with the preceding $a^4$, the coefficient in front of the cosmic scale factor can be eliminated. The velocity shear comes from the varying density perturbation
\begin{align}v_{ij}=-\frac{a}{\delta}\frac{d\delta}{dt}\int\frac{k_ik_j}{k^2}\delta_k e^{-ikx}\frac{d^3k}{(2\pi)^3},
\end{align}
the velocity shear can be decomposed into two parts: the integral part gives a value proportional to the variance, and the part outside the integral gives the time evolution factor, denoted by $g(t)=-\frac{a}{\delta}\frac{d\delta}{dt}$ ~\cite{Heavens:1988},
\begin{align}v_{ij}=g(t)\Tilde{v}_{ij}.
\end{align}

In peak theory~\cite{Bardeen:1985tr,DeLuca:2019buf}, the angular momentum of a black hole is correlated with the height of the density peak. In our framework, we posit that black holes form once the perturbation attains a specific threshold peak value. Primordial black holes are produced at different times through distinct nucleation processes. We argue that, at the corresponding horizon‑entry epoch, the overdense region on the relevant scale can undergo non‑spherical collapse, leading to the formation of spinning primordial black holes.
\begin{figure*}
    \centering
    \includegraphics[width=12cm]{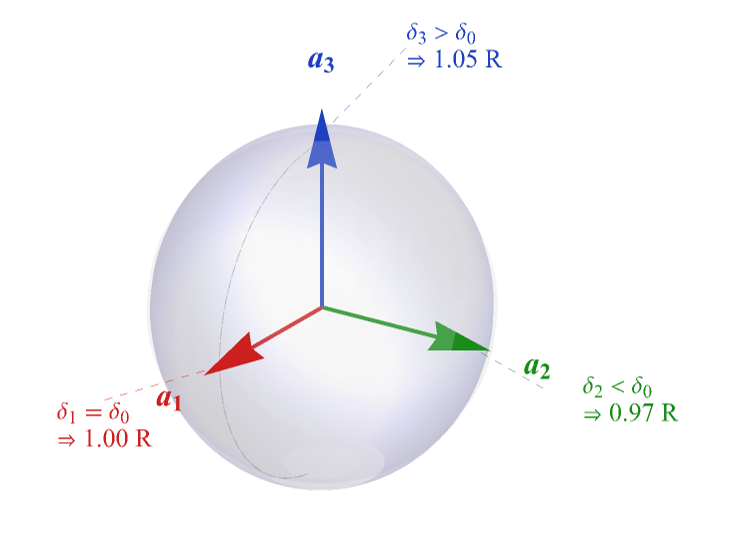}
   \caption{ Geometric origin of non-spherical collapse. The semi-axis $a_i$ of the overdense ellipsoidal region deviates from the background horizon scale $R = 1/(aH)$ according to the local density perturbation $\delta_i$ measured along principal axis $i$. The three directions exhibit independent deviations ($\delta_1 = \delta_0$, $\delta_2 < \delta_0$, $\delta_3 > \delta_0$), yielding $a_1 = R$, $a_2 < R$, and $a_3 >R$, respectively. This ellipsoidal asymmetry, seeded by stochastic bubble nucleation during the first-order phase transition, constitutes the geometric prerequisite for tidal torque generation of primordial black hole spin.}
    \label{}
\end{figure*}

\section{False-vacuum islands}\label{sec:island}
In this chapter, we investigate the generation and evolution of density perturbations induced by a first-order phase transition in the early Universe. The physical origin of these perturbations lies in the stochastic nature of vacuum bubble nucleation: different Hubble patches begin the phase transition at different times, and those with delayed nucleation remain in the false vacuum for longer, thereby storing more vacuum energy than the background. This leads to local overdensities relative to the surrounding radiation bath. Our goal is to quantify this effect by determining the background evolution during the phase transition, computing the density contrast associated with different nucleation histories, and then constructing a statistical description of the resulting perturbation field. In particular, we will calculate the mean density perturbation, its standard deviation, and, more generally, the expectation values of arbitrary functions of the density perturbation.

We begin with the false-vacuum volume fraction ~\cite{Cai:2024nln}
\begin{align}\label{eq:3.1}
F(t)=e^{-\frac{4\pi}{3}\int_{t_i}^{t}dt'\Gamma(t'){a(t')^3}{r(t,t')^3}},\end{align}
where $r(t,t')=\int_{t'}^{t}dt'/a(t')$ is the comoving radius at time $t$ of a vacuum bubble produced at time $t'$. And the $\Gamma$ is nucleation rate of
vacuum bubbles. In this paper we adopt the definition $\Gamma=\Gamma_0e^{\beta t}$ from ~\cite{Cai:2024nln}.We assume that bubbles expand at the speed of light ( 
$c=1$). Prior to the onset of the phase transition, the false‑vacuum fraction equals unity; it then decreases monotonically toward zero as the transition proceeds. The vacuum energy is transferred to the bubble walls and the background plasma. Since both the bubble-wall energy density and the radiation energy density scale as $a^{-4}$ after the phase transition (the wall energy is quickly dissipated into the surrounding plasma), they can be combined into a single effective radiation component for the purpose of background evolution. The nucleation time for the first bubble varies among Hubble patches owing to the probabilistic character of quantum tunneling. Regions that nucleate later retain a higher vacuum energy density. Throughout this work, the phase transition is taken to occur during the radiation‑dominated epoch. 

The evolution of the vacuum energy density and the radiation energy density is determined by the energy conservation equation  with the Friedmann equation.
\begin{align}\frac{d\rho_r}{dt}+4\rho_rH=-\frac{d\rho_v}{dt},\end{align}
\begin{align}H^2=\frac{a'(t)^2}{a(t)^2}=\frac{\rho_r+\rho_v}{3},\end{align}
where $\rho_v = F\Delta V$ is the vacuum energy density. During the radiation‑dominated era, for which $a(t) \propto t^{1/2}$, the false vacuum fraction $F(t)$ can be obtained by substituting this scaling into Eq.~\eqref{eq:3.1}. With $F(t)$ determined, the Friedmann equations can be solved, yielding the evolution of the radiation and vacuum energy density throughout the phase transition. However, this simplified approach can become inaccurate in regions where the vacuum energy density is excessively high, owing to the breakdown of the radiation-domination assumption.

Because of false vacuum fraction $F(t)$ is influenced by the cosmic scale factor $a(t)$.  Since the solution of the Friedmann equations in turn depends on $F(t)$, we are therefore required to solve a coupled system of integro‑differential equations.We can address this by transforming the two-dimensional integro-differential equation into a seven-dimensional system of first-order ordinary differential equations.
To solve the equations, we introduce auxiliary functions: $R(t), v_0(t), v_1(t), v_2(t), v_3(t)$, which are constructed from the integrals present in the original equation.

 The auxiliary function $R(t)$ satisfies
\begin{align}R(t)=\int^t_{t_c}dt'/a(t'),\end{align}
\begin{align}R'(t)=1/a(t),\end{align}
we define $v_i(t), i=0,1,2,3$ as
\begin{align}v_i(t)=\int^t_{t_c}\Gamma(t')a^3(t')R^i(t')dt',\end{align}
\begin{align}v_i'(t)=\Gamma(t)a^3(t)R^i(t),\end{align}
the vacuum bubble radius satisfies
\begin{align}r(t,t')^3=(R(t)-R(t'))^3=R(t)^3-3R^2(t)R(t')+3R(t)R^2(t')-R(t')^3,\end{align}
the false vacuum fraction is $F=e^{I(t)}$, where
\begin{align}I(t)=\frac{4\pi}{3}[R^3(t)v_0(t)-3R^2(t)v_1(t)+3R(t)v_2(t)-v_3(t)],\end{align}
thus, we transform the second-order integro-differential equation into a seven-dimensional system of first-order ODEs:
\begin{align}\frac{a'(t)^2}{a(t)^2}=\rho_r+e^{\frac{4\pi}{3}[R^3(t)v_0(t)-3R^2(t)v_1(t)+3R(t)v_2(t)-v_3(t)]}\Delta V,\end{align}
\begin{align}\frac{d\rho_r}{dt}+4\left(\frac{a'(t)}{a(t)}\right)\rho_r=\frac{de^{\frac{4\pi}{3}[R^3(t)v_0(t)-3R^2(t)v_1(t)+3R(t)v_2(t)-v_3(t)]}}{dt}\Delta V .\end{align}
 
 We can obtain the density evolution $\rho_r(t,t_n)$, $\rho_v(t,t_n)$ for Hubble regions starting the phase transition at different times
 ( regions with different nucleation processes) , where $t_n$ denotes the local phase transition start time. For most regions, $t_n=t_i$, so we consider the evolution of the background density as $\rho_r(t,t_i)$, $\rho_v(t,t_i)$. The density perturbation in delayed phase transition regions is
\begin{align}\delta=\frac{\rho_r(t,t_n)+\rho_v(t,t_n)}{\rho_r(t,t_i)+\rho_v(t,t_i)}-1.\end{align}
\begin{figure}
    \centering
    \includegraphics[width=12cm]{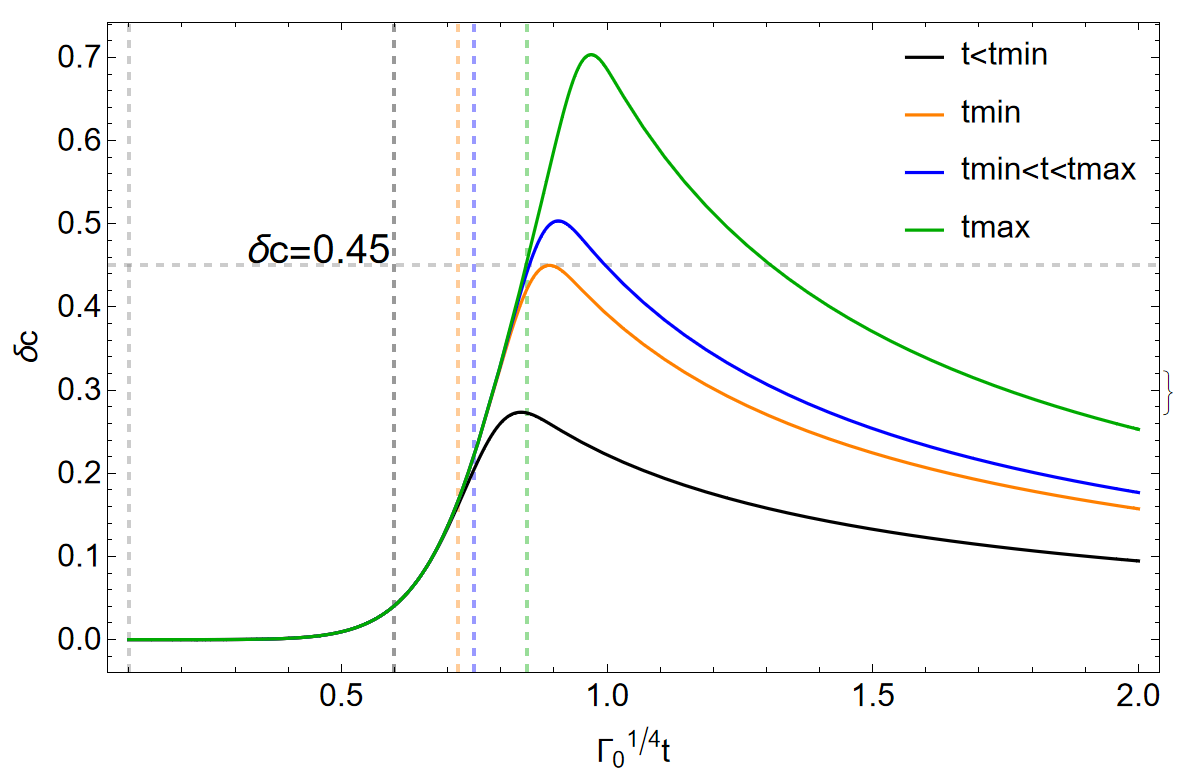}
    \caption{The evolution of density perturbations in Hubble patches with different nucleation processes.
    }
    \label{fig:my_label}
\end{figure}
In figure ~\ref{fig:my_label} , Different colored vertical lines represent the nucleation times of vacuum bubbles in distinct Hubble patches, while the corresponding curves illustrate the evolution of density perturbations within those patches where the phase transition begins at these respective times. The black curve peaks below the primordial black hole  formation threshold $\delta_c=0.45$~\cite{Carr:1974nx,Niemeyer:1997mt}; hence, no PBH are produced. In contrast, the peak of the orange curve exactly meets this threshold. The orange vertical line therefore defines the minimum delayed phase transition time necessary for PBH formation: only regions that begin to transition to the right of this line can produce PBH.When a region starts its transition at the time indicated by the green vertical line, its density excess attains the peak value immediately. This implies that by this time, PBH form irrespective of whether a phase transition occurs. Consequently, the green line represents both the maximum allowable delay for the phase transition and the earliest possible time for PBH formation . Regions that transition between the orange and green lines are capable of forming PBH. Taking the blue line as an example, a region beginning its transition at this time will form a PBH when its density perturbation (the blue curve) intersects the gray horizontal threshold line. Notably, a later phase transition onset leads to an earlier PBH formation time.

Next, we consider the probability for a Hubble volume not to decay until time $t_n$. Due to the definition of nucleation rate $\Gamma$, we find that the probability of a Hubble volume decay at time $t$ to $t+dt$ is $\Gamma(t)V_H(t)dt$ and the probability of a Hubble volume not to decay at this time interval is 
\begin{align}dp(t)=1-\Gamma(t)V_H(t)dt\approx e^{-\Gamma(t)V_H(t)dt}.\end{align}
Consequently, the total survival probability from the initial time $t_i$ until $t_n$ is the product of these infinitesimal probabilities, which yields
\begin{align}P(t_n)=\prod_{t=t_i}^{t=t_n}dp(t)=\mathrm{exp}\left[-\int_{t_i}^{t_n}dt\Gamma(t)V_H(t)\right].\end{align}
We now understand the density perturbation $\delta(t, t_n)$ of a delayed phase transition region with nucleation time $t_n$. When it reaches the critical threshold $\delta_c$, a primordial black hole (PBH) is formed, a moment denoted by $\delta(t_{pbh}, t_n) = \delta_c$. At this point, we consider that the scale entering the horizon $R = 1/(a(t_{pbh}) H(t_{pbh}))$ collapses to form the PBH. The primordial black hole mass is proportional to the mass of the overdense region. Furthermore, from Eq. (3.14), we obtain the probability $P(t_n)$ that a Hubble region remains in the false vacuum  until time $t_n$. The probability for a false vacuum of volume $V$ to transition between $t$ and $t+dt $ is $ \Gamma(t) V(t) dt$, which leads to the probability density function for the nucleation time, $P(t_n) \Gamma(t) V(t) dt$. The ensemble average of $\delta_R(t)$ taken over all independent spherical regions of fixed comoving radius $R$.

\begin{align}
\langle\delta_{R}(t,\mathbf{x})\rangle_{\mathbf{x}}&\equiv\langle\delta_R(t; t_i+\Delta t(\mathbf{x}))\rangle_{\mathbf{x}\in\mathbb{R}^3/R^3}\equiv\langle\delta(t;t_n)\rangle_{t_n}=\int_{t_i}^{U(t)}\delta(t;t')P_R(t')\Gamma(t')V_R(t')\mathrm{d}t',
\end{align}
where in the first congruence, $\mathbb{R}^3/R^3$ represents the quotient set of three-dimensional Euclidean space with respect to scale $R$, dividing the entire space into spheres of comoving radius $R$; the second congruence comes from converting the spatial average into a time average, so we can express the integral over the entire space by integrating over different phase transition start times $t_n$ for different regions. The integral upper limit function is chosen as~\cite{Cai:2024nln}, we define $t_{\mathrm{PBH}}^{\min}$ as the earliest time when any PBH forms, and $t_{\mathrm{PBH}}^{\max}$ the latest. Correspondingly, $t_n^{\max}$ is the latest nucleation time leading to PBH formation at $t_{\mathrm{PBH}}^{\min}$, and $t_n^{\min}$ the earliest nucleation time leading to PBH formation at $t_{\mathrm{PBH}}^{\max}$.

\begin{align}
U(t)=\begin{cases}
t_n^\mathrm{max}, &\quad t<t_n^\mathrm{max}\equiv t_\mathrm{PBH}^\mathrm{min},\\
t_n(t), &\quad t_n^\mathrm{max}\equiv t_\mathrm{PBH}^\mathrm{min}<t<t_\mathrm{PBH}^\mathrm{max},\\
t_n^\mathrm{min}, &\quad t>t_\mathrm{PBH}^\mathrm{max}.
\end{cases}
\end{align}
Below, we detail the core part of our proposed nucleation history summation method for calculating density perturbations, namely the selection of the integral upper limit function $U(t)$: (i) When the considered time is earlier than the earliest primordial black hole formation time,  $t<t_n^\mathrm{max}\equiv t_\mathrm{PBH}^\mathrm{min}$, all regions with comoving scale $R$ contribute to the nucleation history summation integral, so we need to consider all regions that start the phase transition from $t_i$ to $U(t)=t_n^\mathrm{max}$, regardless of whether these regions later form primordial black holes. (ii) When the considered time falls within the time interval,  $t_\mathrm{PBH}^\mathrm{min}<t<t_\mathrm{PBH}^\mathrm{max}$, besides the regions that have already collapsed into primordial black holes, all remaining regions with comoving scale $R$ contribute to the nucleation history summation integral. The regions that form primordial black holes at time $t$ started the phase transition at $U(t)=t_n(t)$, where $t_n(t)$ satisfies $\delta(t; t_n(t), t_i)=\delta_c$. Because the later the delayed phase transition, the earlier the primordial black hole forms, regions that start the phase transition after $t_n(t)$ have already collapsed into primordial black holes before time $t$, so we do not include the contribution from these delayed phase transition regions. Accordingly, we choose the integral upper limit function as $t_n(t)$. (iii) When the considered time is later than the latest primordial black hole formation time, $t>t_\mathrm{PBH}^\mathrm{max}$, all delayed phase transition regions that can form primordial black holes have already collapsed into primordial black holes. Therefore, we only need to consider regions that do not form primordial black holes, which start the phase transition between $t_i$ and $U(t)=t_n^\mathrm{min}$. The above nucleation history summation process can be used to calculate the average value of any function of the density perturbation field as follows,
\begin{align}\label{eq:3.17}
\langle f(\delta(t))\rangle=\int_{t_i}^{U(t)}f(\delta(t;t'))P(t')\Gamma(t')V(t')\mathrm{d}t'.
\end{align}

In the next chapter, we will establish the connection between density perturbations and the semi-axes of ellipsoidal collapse regions. By the nucleation process integration as described above, we will compute the mathematical expectation of the semi-axes, which will then allow us to determine the angular momentum of the overdense regions.

\section{kerr parameter from nucleation process integration }\label{sec:PBH}
During the FOPT process, the asynchronicity of the phase transition leads to inhomogeneities in density perturbations, resulting in non-spherical collapse in overdense regions. Furthermore, the varying density perturbations produce a velocity shear tensor that is misaligned with the inertia tensor of the overdense regions. According to tidal torque theory, this misalignment gives rise to rotating primordial black holes. The scale of the overdense region (the radius of the sphere in spherical collapse) is determined by the scale at which the density perturbation reaches the critical value $\delta_c = 0.45$ upon horizon entry. If density perturbations exist outside this scale along the three principal axes, causing the background density to be perturbed such that $\delta_{x,y,z} = \delta(t_i + \Delta t_{x,y,z}) > \delta(t_i) = \delta_{\text{background}}$, then the semi-major axes along the corresponding principal directions deviate from or exceed the horizon scale $1/aH$. This is the origin of non-spherical symmetry. Since we perform a second-order expansion of the density perturbation peak, the overdense region maintains ellipsoidal symmetry, meaning we only consider the three degrees of freedom of the ellipsoid's principal axes and use the mathematical expectation of the semi-major axes to describe the non-spherical symmetry of the collapse. The actual collapse process can be more asymmetric.
Thus, we can obtain the expectation of the ellipsoidal semi-axes from ~\eqref{eq:2.10}~\eqref{eq:2.11}~\eqref{eq:3.17}
\begin{align}\langle a_i\rangle=\int^{U(t)}_{t_i}\Gamma(t')P(t')V(t')\sqrt{\frac{\delta_{pk}-\delta(t_0)}{a^2H^2(\delta_{pk}-\delta(t'))}}dt',
\end{align}
where $\delta_0=0$ is the density perturbation of the nucleation Hubble patch at time $t_0$, which we treat as the background density perturbation. $\delta_{pk}$ denotes the peak perturbation in the center of the collapsed overdense region. In our work, a primordial black hole is considered to form when the average perturbation of the overdense region reaches the critical threshold $\delta_c=0.45$. Under the second-order approximation, we take $\delta_{pk}=1.12$. Moreover, when evaluating the influence of the surrounding density perturbations on the semi-major axis of the collapsing region, we neglect configurations in which two adjacent Hubble patches both collapse to form primordial black holes. This implies that the upper limit $U(t)$ of our nucleation history integral must not exceed the minimum nucleation time $t_n^{\min}$ that allows primordial black hole formation.

Since $\langle (a_i^2-a_j^2)^2\rangle=\sqrt{2}\sigma_{a^2}$, we have
\begin{align}\sigma_{a^2}=\int^{U(t)}_{t_i}\Gamma(t')P(t')V(t')\left(\sqrt{\frac{\delta_{pk}-\delta(t_0)}{a^2H^2(\delta_{pk}-\delta(t'))}}-\langle a_i\rangle \right)^2dt'.
\end{align}

 \begin{figure*}
    \centering
    \includegraphics[width=12cm]{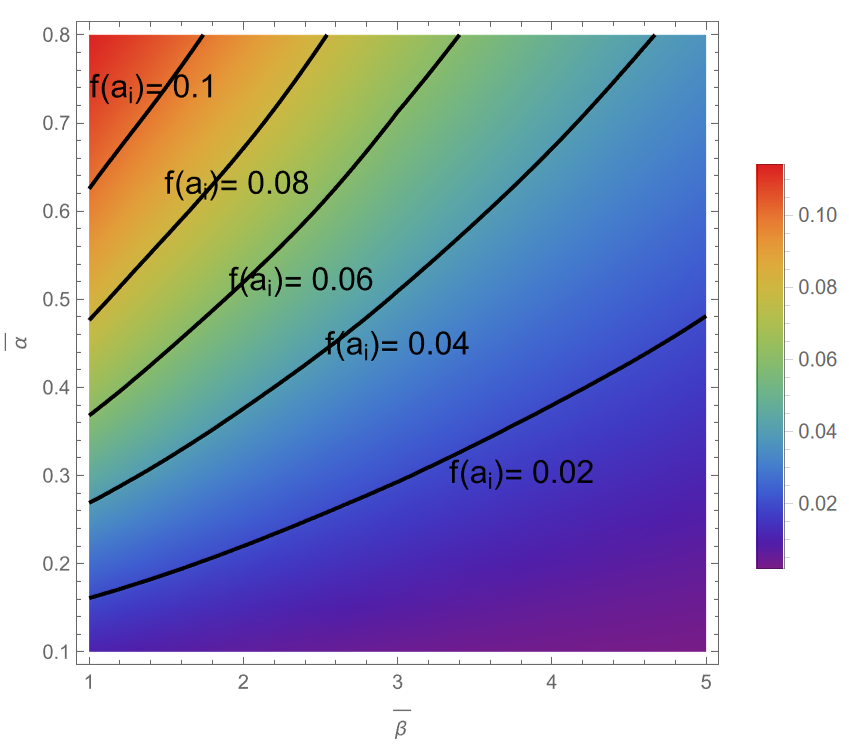}
   \caption{ In the figure 2, we define $f(a_i) = (a_i - R)/R$ to indicate the fractional excess of the semi-major axis of the overdense ellipsoidal region over the horizon scale $R = 1/aH$. It can be observed that a larger phase transition latent heat $\alpha$ and a smaller phase transition strength $\beta$ result in a longer semi-major axis of the ellipsoid relative to the horizon scale.}
    \label{}
\end{figure*}
Therefore, we can obtain the full-space angular momentum expectation from ~\eqref{eq:2.16}
\begin{align} \langle S_i^2\rangle^{1/2}=\frac{16\pi}{45}\langle a_i\rangle^3\sqrt{2}\sigma_{a^2}\langle v_{jk}\rangle\langle\rho_\mathrm{PBH}\rangle=\frac{16\sqrt{2}\pi}{45}\langle a_i\rangle^3\sigma_{a^2}\langle g(\eta)\rangle\frac{1}{\sqrt{15}}\sigma_0\langle\rho_\mathrm{PBH}\rangle.
\end{align}
the semi-axis lengths in the three directions are independent of each other,so $\langle a_ia_ja_k\rangle=\langle a_i\rangle^3$,
the velocity shear can be decomposed into a time-evolving part and a time-independent part. The time-independent part can be related to the variance of the density perturbation
\begin{align}\langle \Tilde{v}_{12}^2\rangle=\int\frac{d^3k}{(2\pi)^3}\frac{d^3k'}{(2\pi)^3}\frac{k_1k_2}{k^2}\frac{k_1'k_2'}{k'^2}\langle\delta_k\delta_k'\rangle e^{i(k+k')x}=\int\frac{d^3k}{(2\pi)^3}\frac{k_1^2k_2^2}{k^4}p_\delta(k)=\frac{1}{15}\sigma_0^2,
\end{align}
\begin{align}\langle \Tilde{v}_{ij}^2\rangle=\frac{1}{15}\sigma_0^2.
\end{align}
the time evolution factor
\begin{align} \langle g(\eta)\rangle=\int^{U(t)}_{t_i}\Gamma(t')P(t')V(t')\frac{\delta'(t')}{\delta(t')}dt'.
\end{align}

Then we define the dimensionless Kerr parameter $a_*=S/M_\mathrm{PBH}^2$ that describes the spin of black holes. The mass of PBH is $M_\mathrm{PBH}=\gamma \frac{4\pi}{3} H_\mathrm{PBH}^{-3}\rho_\mathrm{PBH}$.
\begin{align} \langle a^2\rangle^{1/2}=\sqrt{3}\left(\frac{16\sqrt{2}\pi}{45}\langle a_i\rangle^3\sigma_{a^2}\langle g(\eta)\rangle\frac{1}{\sqrt{15}}\sigma_0\langle\rho_\mathrm{PBH}\rangle\right)/\left(\gamma \frac{4\pi}{3} H_\mathrm{PBH}^{-3}\langle\rho_\mathrm{PBH}\rangle\right)^2.
\end{align}
\begin{figure*}
    \centering
    \includegraphics[width=12cm]{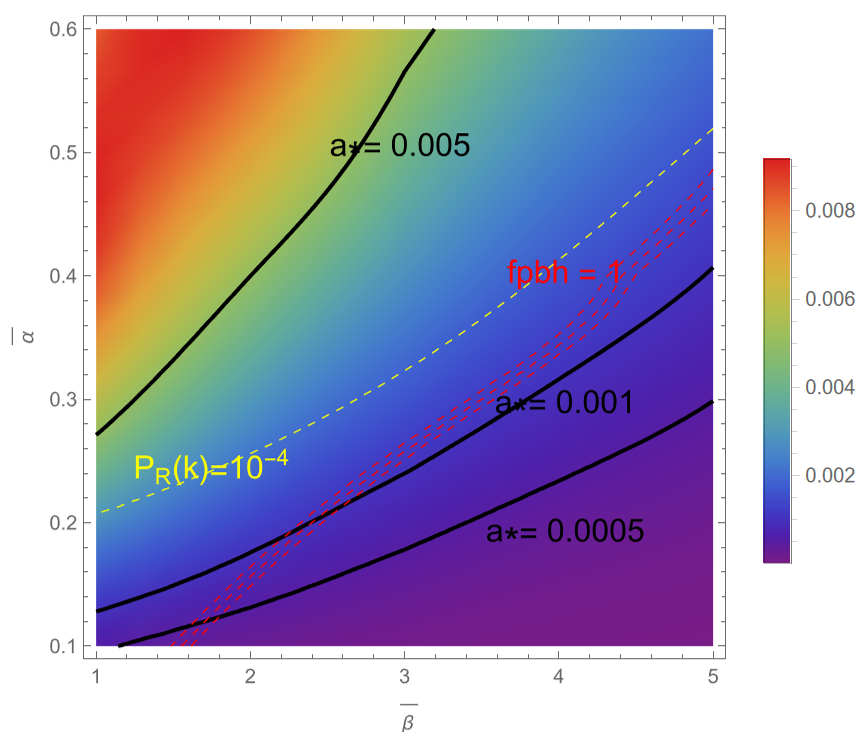}
   \caption{ The figure illustrates the relationship between the Kerr parameter and the phase transition latent heat $\alpha$ and the phase transition duration parameter $\beta$, and also presents the observational constraints on the primordial black hole abundance from $10^{-16}$ sun mass to $10^{-12}$ sun mass, and the curvature perturbation power spectrum at the horizon entry during the formation epoch of primordial black holes, based on the analysis in reference \cite{Cai:2024nln}.}
    \label{}
\end{figure*}

In peak theory,the density fluctuation field $\delta(\mathbf{x})$ is assumed to be a Gaussian random field. To describe the physical state of a density peak, we define a 16-dimensional variable set $V_i$, which includes density contrast $\delta$, its gradient $\xi_i = \nabla_i \delta$, the Hessian matrix $\xi_{ij} = \nabla_i \nabla_j \delta$, and the tidal shear components $v_{ij}$.The probability that variables $V_i$ take specific values is given by the multivariate Gaussian distribution.
\begin{align}f(V_i)d^{16}V_i=\frac{1}{(2\pi)^8M^{1/2}}e^{-\frac{1}{2}(V_i-\langle V_i\rangle)M_{ij}^{-1}(V_j-\langle V_j\rangle)}d^{16}V_i
    \end{align}  
where $M_{ij}$ is the covariance matrix representing the correlations between these spatial derivatives,
\begin{align}M_{ij}=(V_i-\langle V_i\rangle)(V_j-\langle V_j\rangle)
    \end{align}    
We can give the probability distribution of any function composed of the above 16 parameters, thus the Kerr parameter is
 \begin{align}a_*=10^{-2}\sqrt{1-\gamma^2}
    \end{align} 
\begin{align}\gamma=\frac{\sigma_0\sigma_2}{\sigma_1^2}
    \end{align} 
where $\sigma_0$, $\sigma_1$, and $\sigma_2$ are the standard deviation of the density perturbation, the standard deviation of its first derivative, and the standard deviation of its second derivative, respectively.    
The parameter $\gamma$ is the spectral shape parameter,which measures the width of the power spectrum and determines the typical shape of the peaks,$\gamma$ ranges from $0$ to $1$, $\gamma \to 1$ implies highly spherical peaks produced by a narrow power spectrum.

We now summarize the relationship between the Kerr parameter and the phase transition parameters $\alpha$ and $\beta$. It can be seen from the figures that the Kerr parameter increases with increasing $\alpha$ and decreases with increasing $\beta$. At the energy scale corresponding to $\mu$-distortion, $a_*$ is on the order of $10^{-3}$. This result represents a significant enhancement compared to PBHs formed in the standard radiation-dominated era \cite{Chiba:2017rvs,DeLuca:2019buf} and those produced from first-order phase transitions \cite{Banerjee:2023qya}, which typically have $a_* \sim 10^{-5}$, while PBHs formed in a matter-dominated era can achieve Kerr parameters up to $\sim 10^{-1}$. The angular momentum of PBHs follows the same dependence on the phase transition parameters $\alpha$ and $\beta$ as the PBH abundance. For instance, for $\alpha=0.2$, $\beta=2.4$, we have $f_{\mathrm{PBH}}=1$ in the window mass and $a_*\approx 10^{-3}$. Furthermore, the angular momentum of PBHs can be further enhanced through various particle physics and astrophysical mechanisms, such as superradiant instability \cite{Calza:2023rjt} and PBH mergers \cite{Hofmann:2016yih}.

\begin{table}[htbp]
\centering
\caption{\textbf{Comparison of Primordial Black Hole (PBH) Spin Parameters ($a_*$).} This table contrasts the dimensionless spin parameter across different formation mechanisms and evolutionary pathways. }
\label{tab:spin_comparison}
\vspace{2mm} 
\begin{tabular}{|l|c|c|}
\hline
\textbf{Mechanism / Model} & \textbf{Formation Epoch} & \textbf{Spin Parameter $a_*$} \\ \hline
Peak Theory (FOPT) & Radiation-Dominated & $\sim 10^{-5}$ \cite{Banerjee:2023qya} \\ \hline
\hline
\textbf{Nucleation History Integration (FOPT)} & \textbf{Radiation-Dominated} & $\mathbf{\sim 10^{-3}}$ \\ \hline
\hline
Early Matter-Dominated & Matter-Dominated & $\sim 10^{-1}$ \cite{Harada:2017fjm} \\ \hline
Hierarchical Mergers & Late-time Evolution & $\approx 0.7$ \cite{Fishbach:2017dwv} \\ \hline
Gas Accretion & Late-time Evolution & $\approx 0.998$ \cite{Thorne:1974ve,DeLuca:2020bjf} \\ \hline
\end{tabular}
\end{table}

It is crucial to emphasize that the PBH spin magnitude derived in this study likely represents a conservative lower bound. During a first-order phase transition (FOPT), the highly dynamic environment characterized by bubble colli sions, sound waves, and turbulence \cite{Hindmarsh:2015qta,Caprini:2015zlo} generates substantial velocity shear and local vorticity \cite{Cutting:2019zws}. This hydrodynamic injection of angular momentum explicitly breaks the spherical symmetry of the collapsing region, potentially leading to a significantly higher initial spin than predicted by radiation-dominated collapse models. Furthermore, this initial primordial spin can be continuously amplified through cosmological dynamical evolution \cite{DeLuca:2020bjf}. Hierarchical mergers within PBH clusters efficiently convert orbital angular momentum into remnant spin,( typically yielding $a_*=0.68$ for equal-mass non-spinning binaries) \cite{Hofmann:2016yih}, while accretion can further spin up the black holes toward the Thorne limit ($a_*=0.998$) \cite{DeLuca:2020bjf,Thorne:1974ve}. Consequently, the combination of FOPT-driven hydrodynamic shear and subsequent astrophysical processes suggests that the final observable PBH spins could be substantially larger than our estimates. This would yield valuable astronomical observational effects.


\section{Conclusions and discussions}\label{sec:condis}

This paper presents the relationship between the Kerr parameter of primordial black holes and the FOPT parameters $\alpha$ and $\beta$, while also examining the relationship between the abundance of primordial black holes and the FOPT parameters $\alpha$ and $\beta$.  When studying the spin of primordial black holes, we should exclude parameter spaces where $f_{\text{PBH}} > 1$ and subject the power spectrum to astronomical observational tests, including spectral distortions in the cosmic microwave background radiation. Additionally, parameter spaces with a Kerr parameter greater than 1 would form extreme black holes containing naked singularities, whose physical existence remains uncertain. We also impose constraints on $\alpha$, as excessively large phase transition latent heat would violate the assumption of radiation domination.

In peak theory, the angular momentum of black holes is related to the peak height. In our work, we assume that black holes form when perturbations reach a specific peak value. Different nucleation processes form primordial black holes at different times. We propose that, at the moment of horizon entry, overdense regions can undergo non-spherical collapse, thereby forming primordial black holes. In summary, peak theory studies different peaks at the same time, while our work studies the same peak at different times. Moreover, due to the non-Gaussian nature of perturbations generated during the FOPT process, we use a nucleation history integration method to derive the Kerr parameter describing black hole spin without assuming a Gaussian distribution. Compared to the results of peak theory, our approach yields higher angular momentum.



\bibliographystyle{JHEP}
\bibliography{ref}

\end{document}